\documentclass[twocolumn,american,pre]{revtex4}
\usepackage[T1]{fontenc}
\usepackage[latin1]{inputenc}
\usepackage{geometry}
\usepackage{amsmath}
\usepackage{babel}
\usepackage{graphics}
\usepackage{amssymb}

\makeatletter

\providecommand{\LyX}{L\kern-.1667em\lower.25em\hbox{Y}\kern-.125emX\@}

\makeatother
\begin{document}

\newcommand{\epshat}{\hat{\varepsilon }}

\newcommand{\eps}{\varepsilon }

\title{Unstable attractors induce perpetual synchronization and desynchronization}

\author{Marc Timme}

\author{Fred Wolf }

\author{Theo Geisel}

\affiliation{Max-Planck-Institut für Strömungsforschung and Fakultät für Physik,
Universität Göttingen, 37073 Göttingen, Germany.}

\begin{abstract}
Common experience suggests that attracting invariant sets in nonlinear
dynamical systems are generally stable. Contrary to this intuition,
we present a dynamical system, a network of pulse-coupled oscillators,
in which \textit{unstable attractors} arise naturally. From random
initial conditions, groups of synchronized oscillators (clusters)
are formed that send pulses alternately, resulting in a periodic dynamics
of the network. Under the influence of arbitrarily weak noise, this
synchronization is followed by a desynchronization of clusters, a
phenomenon induced by attractors that are unstable. Perpetual synchronization
and desynchronization lead to a switching among attractors. \textbf{}This
is explained by the geometrical fact, that these unstable attractors
are surrounded by basins of attraction of other attractors, whereas
the full measure of their own basin is located remote from the attractor.
Unstable attractors do not only exist in these systems, but moreover
dominate the dynamics for large networks and a wide range of parameters. 
\end{abstract}
\maketitle
\textbf{As attractors determine the long-term behavior of dissipative
dynamical systems, the notion of attractors is central to studies
in many fields of science. According to the mathematical definitions
of an attractor, states that originate in a certain volume of state
space, the basin of attraction, evolve towards the respective attractor.
Since states slightly perturbed from an attractor often stay confined
to its vicinity and finally return to it, attractors are widely considered
to be} \textbf{\textit{stable}}\textbf{. Attracting yet} \textbf{\textit{unstable}}
\textbf{states are consistent with an attractor definition introduced
by Milnor. There is evidence that such Milnor attractors might not
be uncommon in certain systems that exhibit strange invariant sets
with a fractal geometry. In general, however, unstable attractors
seem to be special cases that have to be constructed artificially
by precisely tuning parameters. Contrary to this intuition, we present
an example of a dynamical system, a network of pulse-coupled oscillators,
in which} \textbf{\textit{unstable}} \textbf{attractors with periodic
dynamics arise naturally. Unstable attractors are shown to prevail
for large networks and a wide range of parameters. In the presence
of arbitrarily weak noise, a perpetual synchronization and desynchronization
of groups of oscillators occurs which leads to a switching among different
attractors. }

\section{Introduction}

The concept of attractors is underlying the analysis of many natural
systems as well as the design of artificial systems. For instance,
the computational capabilities of neural networks are controlled by
the attractors of their collective dynamics \cite{Hopfield,Hertz,Maass}.
Consequently, the nature and design of attractors in such systems
constitute a focus of current research interest \cite{BressloffNC,Diesmann,ErnstPRE,ErnstPRL,GerstnerNC,GerstnerPRL,HanselPRL,Mirollo,Senn,Timme,Tsodyks,vanVreeswijk}.
In general, the state space of a nonlinear dynamical system is partitioned
into various basins of attraction from which states evolve towards
the respective attractors. Since states that are slightly perturbed
from an attractor often stay confined to its vicinity and eventually
return to the attractor, attractors are commonly considered to be
\textit{stable} \cite{Eckmann,Guckenheimer,Katok}. 

In this paper we study the dynamics of networks of pulse-coupled oscillators
\cite{ErnstPRE,ErnstPRL,Mirollo}, in which \textit{unstable} attractors
exist and arise naturally as a \textit{collective} phenomenon. Such
models of pulse-coupled oscillators describe e.g.\ synchronization
in spiking neural networks and the dynamics of other natural systems
as diverse as pacemaker cells in the heart, populations of flashing
fireflies, and earthquakes (cf.\ \cite{BressloffNC,Buck,Diesmann,ErnstPRE,ErnstPRL,GerstnerNC,GerstnerPRL,HanselPRL,Herz,Mirollo,Peskin,Senn,Strogatz,Timme,Tsodyks,vanVreeswijk}).
We identify an analytically tractable network exhibiting unstable
attractors. For this network we demonstrate the existence of attractors
that are unstable and located remote from the volume of their own
basins of attraction. Such attracting yet unstable states are consistent
with a definition of attractors introduced by Milnor, which neither
presumes nor implies stability \cite{Milnor}. In certain other systems
such Milnor attractors might not be uncommon if these systems exhibit
strange invariant sets with a fractal geometry \cite{Ashwin,Kaneko,Lai,Sommerer,Yang}.
More generally, however, attractors that are not stable seem to be
special cases that have to be constructed artificially by precisely
tuning parameters. Contrary to this intuition, in the system considered
here, unstable attractors with regular, \textit{periodic} dynamics
are typical in large networks and persist even if the physical model
parameters are varied substantially. The first discovery of the occurrence
and prevalence of unstable attractors in networks of pulse-coupled
oscillators was reported in \cite{Timme}. Here we give a more detailed
analysis of such unstable attractors and explain the observation that
they only occur for excitatory (phase advancing) interactions but
are absent if the interactions are inhibitory (phase retarding).

We argue that dynamical consequences of unstable attractors may persist
in a general class of systems of pulse-coupled units. Such consequences
include an ongoing switching among unstable attractors in the presence
of noise. In systems where the convergence towards an attractor has
a functional role, such as the solution of a computational task by
a neural network \cite{Hertz,Hopfield,Maass}, switching induces a
high degree of flexibility that may provide the system with a unique
advantage compared to multistable systems: In general, it is hard
to leave a stable attractor after convergence, e.g.\ the completion
of a task. With an unstable attractor, however, a small perturbation
is sufficient to leave the attractor and to switch towards another
one.

This paper is organized as follows: In section \ref{sec:Models_PCO}
we introduce a class of network models of pulse-coupled oscillators.
We briefly review earlier results on synchronization phenomena in
such networks within the framework introduced by Mirollo and Strogatz
that is described in detail in section \ref{sec:Mirollo_Strogatz_model}.
In section \ref{sec:Perpertual_syn_desyn} we describe the numerical
observation that, in the presence of noise, trajectories approach
and retreat from periodic orbits by perpetual synchronization and
desynchronization of groups of oscillators. Together with further
numerical investigations, this leads to the hypothesis, that the observed
dynamics is induced by attractors that are unstable. In section \ref{sec:stability_analysis}
we perform an exact stability analysis of a particularly selected
set of attractors. It is followed by an analysis of the dynamics during
a switching transition between two states (Sec.\ \ref{sec:unstable_attractors})
that is further corroborated by a numerical investigation of the structure
of basins of attractions in state space. These results demonstrate
that unstable attractors indeed exist in the class of networks of
pulse-coupled oscillators considered. Section \ref{sec:prevalence}
completes our analysis showing that unstable attractors prevail in
large networks for a wide range of parameters. The paper concludes
in section \ref{sec:conclusions} with (i) a discussion of the dynamical
consequence of switching among unstable attractors in comparison to
smoothly coupled systems, (ii) a brief presentation of preliminary
results for networks exhibiting different interactions or more complex
structures and (iii) an outlook for future investigations.

\section{Models of pulse-coupled oscillators}

\label{sec:Models_PCO}

In studies of synchronization phenomena in networks of pulse-coupled
oscillators, single elements are often modelled as phase oscillators,
assuming that the dynamics of the amplitude of the oscillation is
less important. Thus there is only one relevant dynamical variable,
\( W_{i}, \) that describes the dynamics of an individual oscillator
\( i \). In models of many natural systems, the oscillator variable
\( W_{i} \) represents an analogue of a potential, like the membrane
potential in the case of a current-driven nerve cell. The dynamics
of a network of pulse-coupled oscillators is commonly described by
a system of coupled ordinary differential equations\begin{eqnarray}
\frac{dW_{i}}{dt} & = & A(W_{i})+B(W_{i})S_{i}(t)\label{eq:DE_coupled_oscW} 
\end{eqnarray}
for \( i\in \{1,\ldots ,N\} \) where \( A \) and \( B \) are continuous
functions and \( S_{i} \) represents the interactions within the
network. The pulse-coupling between oscillators is given by \begin{equation}
\label{eq:internal_current}
S_{i}(t)=\sum _{j=1}^{N}\sum _{m=-\infty }^{\infty }\eps _{ij}K_{ij}(t-t_{j,m})
\end{equation}
 where \( \eps _{ij} \) is the strength of the coupling from oscillator
\( j \) to oscillator \( i \) and the response kernels \( K_{ij}(t) \)
have the property that \( K_{ij}(t)\geq 0 \), \( K_{ij}(t)=0 \)
for \( t<0 \), and \( \int _{-\infty }^{\infty }K_{ij}(t)dt=1 \).
The sum includes all times \( t_{j,m} \) at which oscillator \( j \)
reaches a threshold \( W_{\theta }:=1 \) (from below) for the \( m^{\mathrm{th}} \)
time,\begin{equation}
\label{eq:threshold_W}
W_{j}(t_{j,m})\geq W_{\theta }=1\, ,\, \, \, \left. \frac{dW_{j}(t)}{dt}\right| _{t=t_{j,m}}>0.
\end{equation}
If this threshold is reached, \( W_{j} \) is reset to a potential
\begin{equation}
\label{eq:reset_W}
W_{j}(t_{j,m}^{+}):=W_{\mathrm{reset}}=0
\end{equation}
and a signal is generated that is sent to oscillators \( i \). Commonly,
a unit that is reset and sends out a pulse when it reaches a threshold
is said to {}``fire'' at that instant of time. This form of pulse-coupling
idealizes the fact that in diverse biological systems such as populations
of flashing fireflies or networks of spiking neurons in the brain,
units interact by stereotyped short-lasting signals that are generated
as the state of a unit reaches a threshold. Note that the normalizations
\( W_{\theta }=1 \) and \( W_{\mathrm{reset}}=0 \) are made without
loss of generality. Whereas Eqs.~(\ref{eq:DE_coupled_oscW})--(\ref{eq:reset_W})
describe the interaction dynamics without delays between sending and
reception of a pulse, a delay \( \tau >0 \) can easily be included
by a transformation \( K_{ij}(t)\rightarrow K_{ij}(t-\tau ) \) of
the response kernels.

It is often convenient \cite{vanVreeswijk} (and under weak conditions
possible) to transform the variables \( W_{i} \) according to \begin{equation}
\label{eq:VW}
V_{i}(t):=\frac{1}{C}\int _{0}^{W_{i}(t)}B(w)^{-1}dw
\end{equation}
where \begin{equation}
\label{eq:norm_C}
C=\int _{0}^{1}B(w)^{-1}dw
\end{equation}
 such that (\ref{eq:DE_coupled_oscW}) becomes\begin{eqnarray}
\frac{dV_{i}}{dt} & = & \hat{A}(V_{i})+\frac{1}{C}S_{i}(t)\label{eq:DE_coupled_oscV} 
\end{eqnarray}
for \( i\in \{1,\ldots ,N\} \) where \( \hat{A}(V)=C^{-1}A(W(V))/B(W(V)) \)
is determined solving Eq.~(\ref{eq:VW}) for \( W \). If the duration
of the response of a unit \( i \) to an incoming pulse is sufficiently
brief, the kernels \( K_{ij} \) in (\ref{eq:internal_current}) may
be idealized by the Dirac distribution \begin{equation}
\label{eq:K_delta}
K_{ij}(t)=\delta (t)
\end{equation}
 such that the coupling becomes discontinuous.

\begin{figure}
{\centering \resizebox*{6cm}{!}{\includegraphics{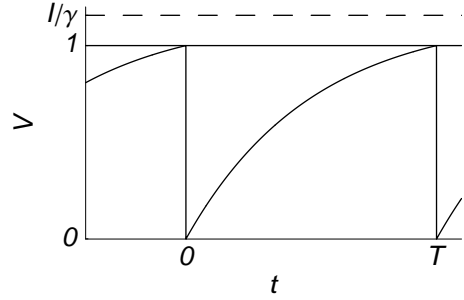}} \par}

\caption{Dynamics of a non-interacting integrate-and-fire oscillator. Without
threshold, the potential \protect\( V\protect \) would converge to
its long-time limit \protect\( I/\gamma \protect \) (dashed line).
Every time the threshold \protect\( V_{\theta }=1\protect \) is reached,
the potential \protect\( V\protect \) is reset to zero and a pulse
is sent. \label{Fig:Peskin_dyn}}
\end{figure}

For the commonly used leaky integrate-and-fire models of oscillators
the free dynamics is given by the linear inhomogeneous differential
equation\begin{equation}
\label{eq:V_linear}
\frac{dV}{dt}=-\gamma V+I
\end{equation}
such that the network dynamics of such oscillators is given by substituting
\( \hat{A}(V_{i})=I-\gamma V_{i} \) in (\ref{eq:DE_coupled_oscV})
where \( I \) is an external current and \( \gamma >0 \) measures
the dissipation in the system. This linear differential equation has
the advantages that the solution is known explicitely and the response
to an additional current can be found by superposition arguments.
For sufficiently large external current, \( I>\gamma  \), the free
(\( S_{i}(t)\equiv 0 \)) dynamics \begin{equation}
\label{eq:V_solution}
\left. \begin{array}{c}
V(t)=I/\gamma \left( 1-e^{-\gamma t}\right) \, \, \, \mathrm{for}\, \, 0<t\leq T\\
V(t+T)=V(t)
\end{array}\right. 
\end{equation}
 is periodic (Fig.~\ref{Fig:Peskin_dyn}) with period \( T=\frac{1}{\gamma }\ln \left( 1-\frac{\gamma }{I}\right) ^{-1} \).

To study the synchronization of pacemaker cells in the heart, Peskin
studied a simple, globally coupled network of such leaky integrate-and-fire
oscillators \cite{Peskin} where the coupling is not delayed, \( \tau =0 \),
and the responses are infinitely fast, \( K_{ij}(t)=\delta (t) \),
excitatory, and homogeneous, \( \eps _{ij}=\eps >0 \). In his 1975
book he conjectured that arbitrary initial conditions converge towards
the fully synchronous state in which all oscillators fire simultaneously.
He gave a proof for \( N=2 \) oscillators assuming that both coupling
strength and dissipation are small, \( \eps \ll 1 \), \( \gamma \ll 1 \). 

In a seminal work of 1990, Mirollo and Strogatz generalized the approach
of Peskin \cite{Mirollo}. Contrary to Peskin, whose analysis was
based on the linearity of the differential equations, Mirollo and
Strogatz's only assumptions were that the free (\( S_{i}(t)\equiv 0 \))
dynamics can be described by a phase-like variable and that the interactions
are mediated by some potential function \( U \) that is a monotonically
increasing and concave down function of this phase \cite{Mirollo}
(for model details see below). This framework includes many cases
for which the associated differential equation is nonlinear such that
its solution may not be known explicitely and superposition arguments
fail. Within this general framework, they proved that, for all \( N \),
almost all initial conditions will ultimately end up in the fully
synchronous state. For systems without dissipation (\( \gamma =0 \)),
equivalent to linear functions \( U \), Senn and Urbanczik \cite{Senn}
proved that the dynamics becomes fully synchronous even if the intrinsic
frequencies and the thresholds of the oscillators are not quite identical.
Like the previous investigators, they treated the case without delay,
\( \tau =0 \), for which the fully synchronous state stays the only
attractor. This periodic orbit is an example of period-one dynamics,
for which every oscillator fires exactly once during one period, and
is the simplest dynamics such a system may exhibit. 

Within the framework introduced by Mirollo and Strogatz, it was, moreover,
shown that the introduction of a delay time \( \tau >0 \), that occurs
ubiquitously in natural systems, changes this situation drastically
\cite{ErnstPRL,ErnstPRE}: With increasing network size, an exponentially
increasing number of attractors coexist. In a large region of parameter
space, these are periodic orbits with period-one dynamics, that exhibit
several groups of synchronized oscillators (clusters), which reach
threshold and send pulses alternately. It was also shown \cite{ErnstPRL,ErnstPRE}
that these kinds of attractors arise not only in excitatorily coupled
networks (\( \eps >0 \)) but also if the oscillators are coupled
inhibitorily (\( \eps <0 \)).

In the following part of this paper, we show that for excitatory coupling,
many of these periodic orbits with period-one dynamics are unstable
attractors. As a consequence, trajectories converge towards these
unstable attractors by synchronization of groups of oscillators into
several clusters but, in the presence of arbitrarily weak noise, diverge
subsequently via desynchronization of clusters.

\section{Mirollo-Strogatz model}

\label{sec:Mirollo_Strogatz_model}

We consider a homogeneous network of \( N \) all-to-all pulse-coupled
oscillators with delayed interactions. A phase-like variable \( \phi _{i}(t)\in (-\infty ,1] \)
specifies the state of each oscillator \( i \) at time \( t \) such
that the difference between the phases of two oscillators quantifies
their degree of synchrony, with identical phases for completely synchronous
oscillators. The free dynamics of oscillator \( i \) is given by
\begin{equation}
\label{eq:phi_dot_1}
d\phi _{i}/dt=1.
\end{equation}
Whenever oscillator \( i \) reaches a threshold \begin{equation}
\label{eq:threshold_phi}
\phi _{i}(t)=1
\end{equation}
the phase is reset to zero \begin{equation}
\label{eq:reset_phi}
\phi _{i}(t^{+})=0
\end{equation}
and a pulse is sent to all other oscillators \( j\neq i \), which
receive this signal after a delay time \( \tau  \). The interactions
are mediated by a function \( U(\phi ) \) specifying a 'potential'
of an oscillator at phase \( \phi . \) The function \( U \) is twice
continuously differentiable, monotonically increasing, \( U'>0 \),
concave (down), \( U''<0 \), and normalized such that \( U(0)=0 \)
and \( U(1)=1 \). 

\begin{figure}
{\centering \resizebox*{6cm}{!}{\includegraphics{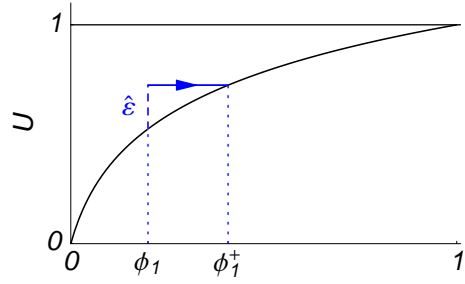}} \par}

\caption{An incoming pulse of strength \protect\( \epshat \protect \) induces
a phase jump \protect\( \phi _{1}^{+}:=\phi _{1}(t^{+})=U^{-1}(U(\phi _{1}(t))+\epshat )=H_{\epshat }(\phi _{1})\protect \)
that depends on the state \protect\( \phi _{1}:=\phi _{1}(t)\protect \)
of the oscillator at time \protect\( t\protect \) of pulse reception.
Due to the monotonicity of \protect\( U\protect \), an excitatory
pulse (\protect\( \epshat >0\protect \)) induces an advancing phase
jump. If the incoming pulse puts the potential above threshold (\protect\( U(\phi _{1})+\epshat >1\protect \)),
the phase is reset to zero (\protect\( \phi _{1}^{+}=0\protect \)).
An inhibitory pulse (\protect\( \epshat <0\protect \)) would induce
a regressing phase jump such that the phase may assume negative values
(not shown). \label{Fig:potential_to_phase} }
\end{figure}

For a general \( U(\phi ) \) we define the transfer function \begin{equation}
\label{eq:H}
H_{\epshat }(\phi )=U^{-1}(U(\phi )+\epshat )
\end{equation}
that represents the response of an oscillator at phase \( \phi  \)
to an incoming subthreshold pulse of strength \( \epshat  \) that
induces an immediate phase jump to \( \phi ^{+}=H_{\epshat }(\phi ) \).
Depending on whether the input \( \epshat  \) is subthreshold, \( U(\phi )+\epshat <1 \),
or suprathreshold, \( U(\phi )+\epshat \geq 1 \), the pulse sent
at time \( t \) (Eq.~(\ref{eq:threshold_phi})) induces a phase
jump after a delay time \( \tau  \) at time \( t'=t+\tau  \) according
to \begin{equation}
\label{eq:phase_jump}
\phi _{j}({t'}^{+})=\left\{ \begin{array}{lll}
H_{\epshat }(\phi _{j}(t')) & \mathrm{if} & U(\phi _{j}(t'))+\epshat <1\\
0 & \mathrm{if} & U(\phi _{j}(t'))+\epshat \geq 1
\end{array}\right. .
\end{equation}
This phase jump (Fig.~\ref{Fig:potential_to_phase}) depends on the
phase \( \phi _{j}(t+\tau ) \) of the receiving oscillator at a time
\( \tau  \) after the signal by oscillator \( i \) has been sent
at time \( t \), the effective coupling strength \( \eps _{ij}=\epshat =\eps /(N-1) \),
and the nonlinear potential \( U \).
 The interactions are either excitatory (\( \varepsilon >0 \)) inducing
advancing phase jumps (Fig.~\ref{Fig:potential_to_phase}) or inhibitory
(\( \varepsilon <0 \)) inducing retarding phase jumps. Note that
in response to the reception of an inhibitory pulse, the phases of
the oscillators may also assume negative values.

The general class of functions \( U \) captures the dynamics of a
variety of systems. In particular, given any differential equation
of the form (\ref{eq:DE_coupled_oscV}) the free (\( S_{i}(t)\equiv 0 \))
dynamics of which has a monotonic periodic solution \( V(t) \) with
period \( T \) and negative curvature, the function \( U \) can
be taken as the scaled solution, \begin{equation}
\label{eq:UV}
U(\phi ):=V(\phi T).
\end{equation}
By this transformation, the general class of pulse-coupled oscillators
(Eqs.\ (\ref{eq:DE_coupled_oscW})--(\ref{eq:reset_W})) with infinitely
fast response (\ref{eq:K_delta}) is mapped onto a normalized phase
description (Eqs.\ (\ref{eq:phi_dot_1})--(\ref{eq:phase_jump})).
For instance, the standard leaky integrate-and-fire oscillator (\ref{eq:V_linear})
with solution (\ref{eq:V_solution}) leads to

\begin{equation}
\label{eq:U_IF}
U_{\mathrm{IF}}(\phi )=\frac{I}{\gamma }(1-e^{-\gamma \phi T})=\frac{I}{\gamma }\left( 1-\left( 1-\frac{\gamma }{I}\right) ^{\phi }\right) .
\end{equation}
 Another example is given by the conductance-based threshold model
of a neuron \cite{Johnston}, in which \( A(W)=\gamma (W_{\mathrm{eq}}-W) \)
and \( B(W)=g(W_{\mathrm{s}}-W) \) with equilibrium potential \( W_{\mathrm{eq}}>1 \),
membrane time constant \( \gamma >0 \), synaptic reversal potential
\( W_{\mathrm{s}} \), and conductivity \( g>0 \). After a transformation
of variables \( V=V(W) \) according to (\ref{eq:VW}) and a scaling
(\ref{eq:UV}) we obtain \begin{equation}
\label{eq:U_conductance_based}
U_{\mathrm{CB}}(\phi )=\frac{\ln \left( 1+W_{\mathrm{eq}}/W_{\mathrm{s}}\left[ \left( 1-W_{\mathrm{eq}}^{-1}\right) ^{\phi }-1\right] \right) }{\ln \left( 1-W_{\mathrm{s}}^{-1}\right) }.
\end{equation}
 The resulting potential function \( U \) is always monotonically
increasing and, for a wide range of biologically reasonable values
of \( W_{\mathrm{s}} \) and \( W_{\mathrm{eq}} \), also concave
down. 

For all numerical studies and simulations presented in this paper,
we use the functional form \begin{equation}
\label{eq:U_b}
U(\phi )=U_{b}(\phi )=b^{-1}\ln \left( 1+(e^{b}-1)\phi \right) 
\end{equation}
that results in an affine transfer function \( H_{\epshat }(\phi )=e^{\epshat b}\phi +\mathrm{const} \),
that was utilized also in the original work of Mirollo and Strogatz
\cite{Mirollo} as well as in the later investigation of the influence
of delay \cite{ErnstPRL,ErnstPRE}. \foreignlanguage{english}{}

Compared to systems of nonlinear differential equations, the Mirollo-Strogatz
approach has the additional advantage, that numerical calculations
can be performed exactly on an event-by-event basis. Given the state
of the system at time \( t \), defined by a phase vector \( \boldsymbol {\phi }(t) \)
and a list of signals sent together with their sending times, the
dynamics can be computed numerically by iterating between two kinds
of events (in an appropriate order):

\begin{enumerate}
\item If the next event is the reception of a signal after a time \( \Delta t_{1} \),
shift all phases \( \phi _{i}(t) \) by this amount and apply the
map \( H_{\epshat } \) according to Eq. (\ref{eq:phase_jump}). If
the received signal is subthreshold for oscillator \( j \), its resulting
phase is given by \begin{equation}
\label{eq:num_update1_receive}
\phi _{j}(t+\Delta t_{1})=H_{\epshat }(\phi _{j}(t)+\Delta t_{1})\, ,
\end{equation}
 if the signal is suprathreshold for oscillator \( j' \),~ \( \phi _{j'}(t)+\Delta t_{1}\geq 1 \),
its phase is reset to zero,\begin{equation}
\label{eq:num_update1_reset}
\phi _{j'}(t+\Delta t_{1})=0.0\, ,
\end{equation}
 and a signal is generated, i.e. its sending time \( t+\Delta t_{1} \)
and the set of sending oscillators \( j' \) is stored. If \( k>1 \)
signals have been simultaneously sent by synchronized oscillators
their simultaneous arrival can simply be numerically realized by an
enlarged coupling strength \( (k-1)\epshat  \) or \( k\epshat  \),
depending on whether or not the considered receiving oscillator has
sent one of the \( k \) signals.
\item If the next event is that one or more oscillators reach threshold
after time \( \Delta t_{2}=1-\max _{i}\phi _{i}(t) \) (without receiving
an additional input signal after a time \( \Delta t_{1}<\Delta t_{2} \)),
those oscillators \( j \) are reset to zero,\begin{equation}
\label{eq:num_update2_reset}
\phi _{j}(t+\Delta t_{2})=0.0\, ,
\end{equation}
and a signal is generated (see event 1.) whereas the phases of all
other oscillators \( j' \) are just translated in time according
to\begin{equation}
\label{eq:num_update2_shift}
\phi _{j'}(t+\Delta t_{2})=\phi _{j'}(t)+\Delta t_{2}\, .
\end{equation}
 
\end{enumerate}
These are the only two possible events that change the state of the
system, such that the simulation time is increased by \( \Delta t_{1} \)
or \( \Delta t_{2} \), depending on the kind of event: To determine
the state at the next time step after time \( t \) numerically, one
first finds the minimum of the time \( \Delta t_{1} \) after which
the next signal would arrive and the time \( \Delta t_{2} \) after
which the next phase would cross threshold without an additional incoming
signal (cf.\ Eq.\ (\ref{eq:phase_jump})). Thus, time steps smaller
than \( \Delta t_{\mathrm{MS}}=\min \{\Delta t_{1}\, ,\Delta t_{2}\} \)
do not need to be considered. Note that \( \Delta t_{\mathrm{MS}} \)
depends on the state of the system such that it changes with time
\( t \). 

This event-based numerics can be exploited for every choice of \( U(\phi ) \),
and is not restricted to \( U_{\mathrm{IF}}(\phi ) \) derived from
the linear differential equation (\ref{eq:V_linear}), for which one
may as well exactly numerically integrate the original differential
equation due to linear superposition (cf.\ \cite{Rotter,Hansel_Num}). 

Hence, in contrast to standard numerical integration of nonlinear
differential equations, for which the time steps \( \Delta t_{\mathrm{DE}} \)
have to be taken sufficiently small and the numerics is only approximate,
the above numerical algorithm for the Mirollo-Strogatz model is exact.
It is also fast, if oscillators constitute synchronized groups, because
then typical time steps are long, \( \Delta t_{\mathrm{MS}}\gg \Delta t_{\mathrm{DE}} \).

\section{Perpetual synchronization and Desynchronization}

\label{sec:Perpertual_syn_desyn}For such pulse-coupled systems, periodic
orbits with groups of synchronized units constitute relevant attractors
\cite{BressloffNC,Buck,Diesmann,ErnstPRE,ErnstPRL,GerstnerNC,GerstnerPRL,Herz,Mirollo,Peskin,Senn,Strogatz,Tsodyks,vanVreeswijk}.
As mentioned above, for the networks of Mirollo-Strogatz oscillators
with delayed interactions (\( \tau >0 \)) considered here, many different
cluster-state attractors with several synchronized groups of oscillators
(clusters) coexist \cite{ErnstPRE,ErnstPRL}. After an initial transient,
these networks settle down onto such a periodic orbit that displays
period-one dynamics with clusters firing successively within each
period. Thus, in particular, phase-differences between oscillators
are constant at all times when a reference oscillator is reset. 
\begin{figure*}
\resizebox*{15.5cm}{!}{\includegraphics{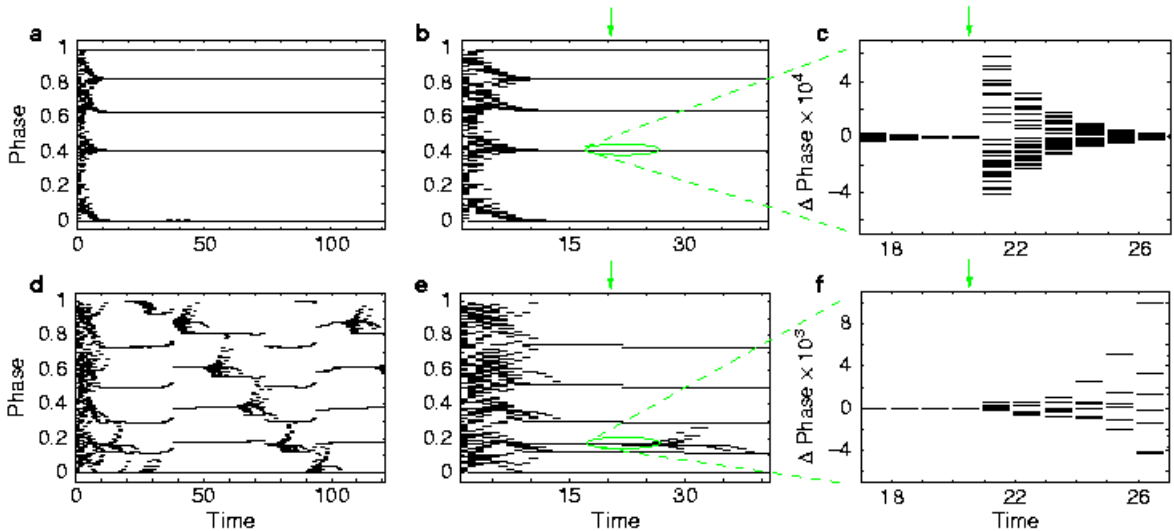}}

\caption{Phase dynamics of a large network (\protect\( N=100\protect \),
\protect\( \tau =0.15\protect \)). Phases of all oscillators are
plotted whenever a reference oscillator has been reset such that the
axis labeled 'Time' is discrete and nonlinear. (a,b,c) Inhibitory
coupling (\protect\( \eps =-0.2\protect \)). (d,e,f) Excitatory coupling
(\protect\( \eps =0.2\protect \)). (a,d) Dynamics with noise (noise
level \protect\( \eta =10^{-3}\protect \)). (b,e) Deterministic dynamics
in response to a single phase perturbation (arrow, perturbation strength
\protect\( \sigma =10^{-3}\protect \)); (b) for inhibitory coupling
the system returns to the attractor; (e) for excitatory coupling the
system switches from a six-cluster to a five-cluster state. (c,f)
Phase differences from the average phase of one cluster (shown in
b and e, respectively) in response to the perturbation. \label{Fig:noisy_phase_dynamics}}
\end{figure*}

Whereas the noise-free dynamics is seemingly similar for both kinds
of coupling, in the presence of noise the collective behavior of excitatorily
coupled oscillators strongly differs from that of oscillators coupled
inhibitorily. For inhibitory coupling all cluster-state attractors
are stable against small perturbations. Under the influence of sufficiently
weak noise the system stays near some periodic orbit that has been
reached after a transient from a random initial state (Fig.~\ref{Fig:noisy_phase_dynamics}a).
The dynamics after a small perturbation in an otherwise noiseless
system confirms this stability property: Perturbations to all clusters
decay exponentially and the original attractor is approached again
(Fig.~\ref{Fig:noisy_phase_dynamics}b,c). 

In contradistinction, for excitatory coupling, we find (cf.\ Fig.~\ref{Fig:noisy_phase_dynamics}d)
that, although the system converges towards a periodic orbit from
random initial states, small noise is often sufficient to drive the
system away from that attractor such that successive switching towards
different attractors occurs. In principle, this dynamics might be
due to stable attractors located close to the boundaries of their
basins of attraction, such that noise drives the trajectory into a
neighboring basin. If this explanation were correct, ever smaller
perturbations off the attractor would lead to an ever lower probability
of leaving its basin. In an otherwise noiseless system we tested this
possibility by applying instantaneous, uniformly distributed, independent
random perturbations \( \delta _{i}\in [0,\sigma ] \) of gradually
decreasing strengths (down to \( \sigma =10^{-14} \)) to the phases
\( \phi _{i} \) of all oscillators \( i \) after the system had
settled down to an attractor (see Fig.~\ref{Fig:noisy_phase_dynamics}e,f).
Even for the weakest perturbations applied, \textit{none} of the perturbed
states returned to the attractor, but all trajectories separated from
the original attractor. We thus hypothesized that the persistent switching
dynamics (Fig.\ \ref{Fig:noisy_phase_dynamics}d) is due to attractors
that are unstable.

\section{Stability analysis}

\label{sec:stability_analysis}
\begin{figure}
{\centering {\small \resizebox*{7.5cm}{!}{\includegraphics{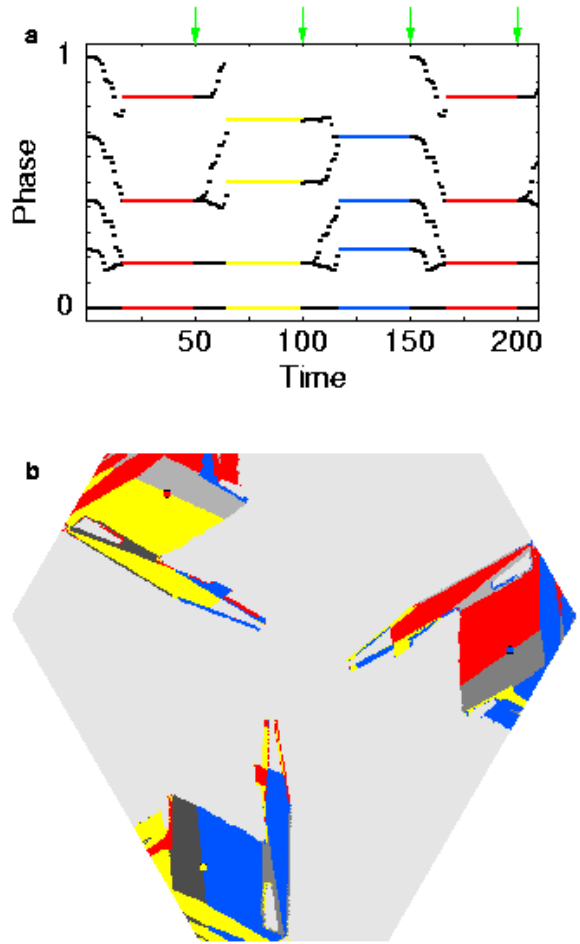}} } \par}

\caption{Small network (\protect\( N=6\protect \), \protect\( \eps =0.2\protect \),
\protect\( \tau =0.15\protect \)) exhibiting unstable attractors.
(a) Noise-free phase dynamics in response to instantaneous perturbations
of magnitude \protect\( \sigma =10^{-3}\protect \) (arrows). When
an attractor is reached, the phase configuration specifying the current
cluster state attractor is marked in color. The perturbations induce
a split-up of clusters and a divergence from the attractor such that
the network reaches different attractors successively. For the realization
of perturbations shown, the attractors marked in red (cluster occupation
list \protect\( [1,2,2,1]\protect \)), yellow (\protect\( [2,2,1,1]\protect \)),
and blue (\protect\( [2,1,1,2]\protect \)) are visited cyclically.
For each of the three attractors there exists another attractor with
the same cluster occupation list but with phase values of the two
clusters with only one oscillator interchanged. (b) Basin structure
of the three attractors color-marked in (a) in two-dimensional planar
section through six-dimensional state space. The planar section is
defined by one point on each of the three periodic orbit attractors,
represented by small red, yellow, and blue disks, respectively. Basins
of the three attractors are marked in the same colors. The part of
state space within this section is of hexagonal shape. Darker gray
areas are basins of the three other permutation-related attractors
(cf.\ (a)) that are located outside the section shown. Lightest gray
marks the union of the basins of all other attractors. \label{Fig:basin_N=3D6} }
\end{figure}

In order to verify this hypothesis
directly, we analyze a small network of \( N=6 \) excitatorily coupled
oscillators for which instantaneous perturbations lead to a similar
switching among attractors. We fix parameters to \( \eps =0.2 \),
\( \tau =0.15 \) and the potential function \( U(\phi )=U_{3}(\phi ) \)
according to Eq.~(\ref{eq:U_b}). At these parameters the network
exhibits a set of periodic orbits with period-one dynamics that are
related by a permutation of phases in such a way, that the system
may switch among them (cf.\ Fig.~\ref{Fig:basin_N=3D6}a, points
on the periodic orbits marked in red, yellow, blue). These orbits
are structurally stable in a neighborhood of these parameters and
the function \( U \). Each attractor can be characterized by a list
of cluster occupation numbers giving the number of oscillators \( n_{k} \)
in the \( k^{\mathrm{th}} \) cluster, counted in the order of increasing
phases at times after a reference oscillator (\( i=1 \)) has been
reset. For instance, the attractor marked in yellow is characterized
by the occupation list \( [2,2,1,1] \) such that the transitions
among the attractors (Fig.~\ref{Fig:basin_N=3D6}a) are described
by the sequence (\( [1,2,2,1] \) (red) \( \rightarrow  \) \( [2,2,1,1] \)
(yellow) \( \rightarrow  \) \( [2,1,1,2] \) (blue) \( \rightarrow  \)
\( [1,2,2,1] \) (red)) for the particular set of perturbations applied.
For each of these three attractors there exists another permutation-related
attractor with the same occupation list but with the phase values
of the two clusters containing only one oscillator interchanged. It
turns out that, depending on the perturbation, transitions from one
attractor occur to one of only two other attractors that are related
through the latter permutation. Thus, once the system has settled
down to one of the attractors shown, it may switch within a set of
six periodic orbit attractors in response to sufficiently small perturbations.

Due to their permutation-equivalence these orbits have identical stability
properties. The state of the network at time \( t \) is specified
by \( \boldsymbol {\phi }(t)=(\phi _{1}(t),\ldots ,\phi _{6}(t))^{\mathsf{T}} \),
such that the orbit marked in yellow in Fig.~\ref{Fig:basin_N=3D6}a
is defined by the initial condition \begin{equation}
\label{eq:syn}
\boldsymbol {\phi }(0)=(0,\, 0,\, A,\, A,\, B,\, C)^{\mathsf{T}}
\end{equation}
 where 

\begin{eqnarray}
A & = & H_{\epshat }(\tau )\label{eq:A} \\
B & = & H_{2\epshat }(1+2\tau -a_{4})\label{eq:B} \\
C & = & H_{2\epshat }(H_{\epshat }(2\tau )+1+\tau -a_{4})\label{eq:C} 
\end{eqnarray}
and recursively defined \begin{equation}
\label{eq:a_i}
a_{i}=U^{-1}(k_{i}\epshat +U(\tau +a_{i-1}))
\end{equation}
 for \( i\in \{1,\ldots ,4\} \), \( a_{0}=0 \), and \( k_{1}=k_{3}=k_{4}=1 \),
\( k_{2}=2 \). Here the origin of time, \( t=0 \), was chosen such
that oscillators \( 1 \) and \( 2 \) have just sent a signal and
have been reset. Moreover, at \( t=0 \) only these two signals (and
no others) have been sent but not yet received. The numerical values
for the particular parameters considered, \( A\approx 0.176 \), \( B\approx 0.499, \)
\( C\approx 0.747 \), can be identified in Fig.~\ref{Fig:basin_N=3D6}a
(orbit marked in yellow). This orbit indeed is periodic, \begin{equation}
\label{eq:phi(0)}
\boldsymbol {\phi }(T)=\boldsymbol {\phi }(0),
\end{equation}
and, in particular, period-one, such that each oscillator fires exactly
once during one period \( T \).

To perform a stability analysis, we define a Poincare map by choosing
oscillator \( i=1 \) as a reference: Let \begin{equation}
\label{eq:phi_ni}
\phi _{n,i}:=\phi _{i}(t_{n}^{+})
\end{equation}
 be the perturbed phases of the oscillators \( i \) at times \( t_{n}>0 \),
\( n\in \mathbb {N} \), just after the resets of oscillator \( 1 \),
\begin{equation}
\label{eq:Poincare_plane}
\phi _{1}(t_{n}^{+})\equiv 0.
\end{equation}
 Thus the five-dimensional vector \begin{equation}
\label{eq:PO'}
\boldsymbol {\delta _{n}}=\boldsymbol {\phi _{n}}-(0,\, A,\, A,\, B,\, C')^{\mathsf{T}}
\end{equation}
 defines the perturbations \( \delta _{n,i} \) for \( i\in \{2,\ldots ,6\} \)
where we choose \( 0<\delta _{n,2} \) and \( \delta _{n,3}<\delta _{n,4} \).
Here\begin{equation}
\label{eq:C_prime}
C'=H_{2\epshat }(H_{\epshat }(2\tau +H_{\epshat }(0))+1+\tau -a_{4})
\end{equation}
 that numerically yields \( C'\approx 0.756. \) Because after a general
perturbation clusters are split up such that in particular \( \phi _{1}<\phi _{2} \),
oscillator \( 6 \) receives the (later) signal from oscillator \( 1 \)
only after it is reset by the (earlier) signal from oscillator \( 2 \)
resulting in \( C'\neq C \). Thus, the original orbit is super-unstable,
i.e. has infinite expansion rate and the stability analysis is performed
for the orbit obtained in the limit \( \boldsymbol {\delta }_{n}\rightarrow \mathbf {0} \).
It is important to note that unstable attractors also exist that do
not posses such a {}``partner orbit'' but have a finite expansion
rate themselves. For convenience, we here continue considering the
above set of orbits that allow a straightforward explanation of the
phenomenon of unstable attraction. Following the dynamics, the five-dimensional
Poincare map is given by \begin{equation}
\label{eq:Poincare}
\boldsymbol {\delta }_{n+1}=\boldsymbol {F}(\boldsymbol {\delta }_{n}),
\end{equation}
 where \( \boldsymbol {F} \) is defined by\begin{equation}
\label{eq:F_i}
\left. \begin{array}{lll}
F_{2}(\boldsymbol {\delta }) & = & 0\\
F_{3}(\boldsymbol {\delta }) & = & L_{4}-L_{3}+H_{\epshat }(\tau -L_{4}+L_{3})-A\\
F_{4}(\boldsymbol {\delta }) & = & H_{\epshat }(\tau +L_{4}-L_{3})-A\\
F_{5}(\boldsymbol {\delta }) & = & H_{\epshat }(H_{\epshat }(1+2\tau -L_{4})+L_{4}-L_{3})-B\\
F_{6}(\boldsymbol {\delta }) & = & H_{\epshat }(L_{4}-L_{3}+H_{\epshat }(\tau +1-L_{3}+\\
 &  & \, \, \, H_{\epshat }(2\tau -\delta _{2}+H_{\epshat }(\delta _{2}))))-C'
\end{array}\right. 
\end{equation}
with the abbreviations \begin{equation}
\label{eq:L_i}
\left. \begin{array}{ccc}
L_{i}:=L_{i}(\boldsymbol {\delta }) & = & H_{\epshat }(\tau +H_{\epshat }(\tau -\delta _{2}+H_{\epshat }(\delta _{2}+\\
 &  & \, \, \, H_{\epshat }(\tau +\delta _{i}-\delta _{2}+H_{\epshat }(\tau )))))
\end{array}\right. 
\end{equation}
 for \( i\in \{3,4\} \). Here the difference \( L_{4}-L_{3} \) in
Eqs.~(\ref{eq:F_i}) is of order \( L_{4}-L_{3}=\mathcal{O}(\left\Vert \boldsymbol {\delta }\right\Vert ) \).

\begin{figure}
{\centering \resizebox*{6cm}{!}{\includegraphics{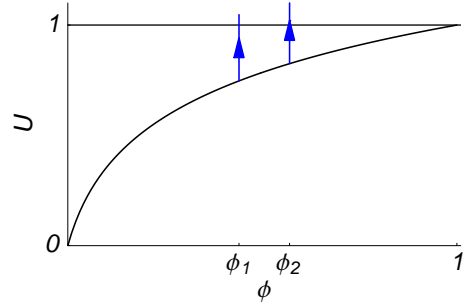}} \par}

\caption{{\small Simultaneous suprathreshold excitatory input synchronizes
immediately due to the reset at threshold. Sufficiently small but
positive phase differences \protect\( \left| \phi _{2}(t)-\phi _{1}(t)\right| >0\protect \)
are reduced to zero such that \protect\( \phi _{2}(t^{+})=\phi _{1}(t^{+})=0\protect \).
This is the mechanism of dimensional reduction of effective state
space, by which attractors can be built.\label{Fig:excsyn}}}
\end{figure}

The linearized dynamics \begin{equation}
\label{eq:M_delta}
\boldsymbol {\delta }_{n+1}\doteq M\boldsymbol {\delta }_{n}
\end{equation}
of a slightly perturbed state with split-up clusters is described
by the Jacobian matrix \begin{equation}
\label{eq:M}
M=\left. \frac{\partial \boldsymbol {F}(\boldsymbol {\delta })}{\partial \boldsymbol {\delta }}\right| _{\boldsymbol {\delta }=\mathbf {0}}=\left( \begin{array}{ccccc}
0 & 0 & 0 & 0 & 0\\
0 & \alpha  & -\alpha  & 0 & 0\\
0 & -\beta  & \beta  & 0 & 0\\
* & * & * & 0 & 0\\
* & * & * & 0 & 0
\end{array}\right) 
\end{equation}
where \( \alpha ,\beta >0 \) and \( * \) denote nonzero real numbers.
It has four zero eigenvalues \begin{equation}
\label{eq:lambda_zero}
\lambda _{i}=0\, \, \, \mathrm{for}\, \, \, i\in \{1,2,3,4\}
\end{equation}
 which imply that a six-dimensional state-space accessed by the random
perturbation is contracted onto a two-dimensional manifold. This reflects
the fact that suprathreshold input received simultaneously by two
or more oscillators (Fig.~\ref{Fig:excsyn}) leads to a simultaneous
reset and thus a synchronization of these oscillators independent
of their precise phases. If a single oscillator is reset by a suprathreshold
signal, it instantaneously exhibits a precise lag in firing time \( \Delta t=\tau  \)
compared to the oscillator that has sent this signal. 

In particular, the zero eigenvalues (Eq.~(\ref{eq:lambda_zero}))
reflect the following contracting dynamics: (i) Perturbations of phases
\( \delta _{n,5}\neq 0 \) or \( \delta _{n,6}\neq 0 \) are restored
immediately by suprathreshold input pulses received from oscillators
\( j=6 \) or \( j=1 \), respectively. This gives rise to the eigenvalues
\( \lambda _{1}=\lambda _{2}=0 \) corresponding to the eigenvectors
\( v_{1}\propto (0,0,0,1,0)^{\mathsf{T}} \) and \( v_{2}\propto (0,0,0,0,1)^{\mathsf{T}} \).
(ii) A splitting of the first cluster, \( \delta _{n,2}>0\equiv \delta _{n,1}\,  \),
corresponding to the vector \( v_{3}\propto (1,0,0,0,0)^{\mathsf{T}} \),
is restored after one period due to one suprathreshold input pulse
from oscillator \( j=4 \). At the same time, however, this splitting
induces a perturbation of oscillators \( i=5 \) and \( i=6 \) such
that \( \delta _{n+1,5}\neq 0 \) and \( \delta _{n+1,6}\neq 0 \)
which is restored according to (i) in the subsequent period. Taken
together, this accounts for the eigenvalue \( \lambda _{3}=0 \).
(iii) As long as \( \delta _{n,3}=\delta _{n,4}\,  \), a perturbation
vector is mapped onto the subspace spanned by \( v_{1} \) and \( v_{2} \)
(see (i)) within one period and is then mapped onto zero during the
next period. As a result, the eigenvalue \( \lambda _{4}=0 \) corresponds
to the direction \( v_{4}\propto (0,1,1,0,0)^{\mathsf{T}} \). In
addition to this analysis, a stability analysis for the subset of
states with the cluster \( \phi _{3}(t)\equiv \phi _{4}(t) \) kept
synchronized, results in superstable directions only, as expected.

\begin{figure}
{\centering \resizebox*{6cm}{!}{\includegraphics{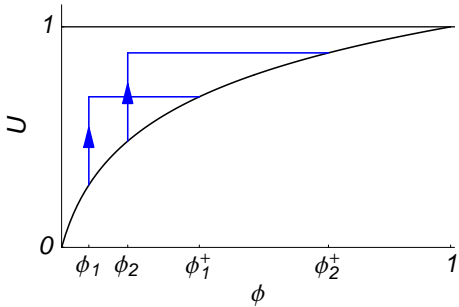}} \par}

\caption{Simultaneous subthreshold excitatory input desynchronizes due to
the concavity of \protect\( U\protect \). Small phase differences
{\small \protect\( \left| \phi _{2}(t)-\phi _{1}(t)\right| >0\protect \)}
are increased, {\small \protect\( \left| \phi _{2}(t^{+})-\phi _{1}(t^{+})\right| >\left| \phi _{2}(t)-\phi _{1}(t)\right| \protect \),}
providing the mechanism that creates an instability. \label{Fig:excdes}}
\end{figure}

In contradistinction to this contracting dynamics, the concavity of
\( U \) implies that simultaneous subthreshold input to two or more
oscillators leads to an increase of their phase differences, i.e.\
a desynchronization of oscillators with similar phases (Fig.\ \ref{Fig:excdes}). 
For the orbits considered here, this is reflected by the only non-zero
eigenvalue\begin{equation}
\label{eq:positive_eigenvalue}
\lambda _{5}=\frac{(2U'(c_{0})-U'(a_{1}))U'(c_{1})U'(c_{2})U'(c_{3})}{U'(a_{1})U'(a_{2})U'(a_{3})U'(a_{4})}>1
\end{equation}
where \begin{equation}
\label{eq:c_i}
c_{i}=\tau +a_{i}
\end{equation}
 for \( i\in \{0,1,2,3\} \) and the \( a_{i} \) are defined in Eq.~(\ref{eq:a_i}).
Because \( c_{i}>a_{i}>c_{i-1} \) for all \( i \) and \( U'>0 \),
\( U''<0 \), this eigenvalue is larger than one, i.e.\ the periodic
orbit is linearly \textit{unstable}. This eigenvalue corresponds to
a split-up of the cluster composed of the oscillators \( i=3 \) and
\( i=4 \). Because the Jacobian (\ref{eq:M}) is not symmetric, the
eigenvectors are not orthogonal such that the corresponding eigenvector
is not \( v\propto (0,1,-1,0,0)^{\mathsf{T}} \) but has a component
in this direction. If there is no homoclinic connection, this instability
implies that such an attractor is not surrounded by a positive volume
of its own basin of attraction, but is located at a distance from
it: Thus, every random perturbation to such an attractor state --
no matter how small -- leads to a switching towards a different attractor.

\section{Unstable Attractors}

\label{sec:unstable_attractors}

Furthermore, this periodic orbit indeed is an \textit{attractor:}
According to the stability analysis, after two firings of the reference
oscillator, a trajectory perturbed off a periodic orbit (e.g. the
one marked in red in Fig.~\ref{Fig:basin_N=3D6}a, which is permutation-equivalent
to the yellow one) is mapped onto a two-dimensional manifold, re-synchronizing
one cluster. The trajectory then evolves towards a neighborhood of
another attractor (here: the yellow one) in a lower dimensional effective
state space without further dimensional reduction. Here, forming the
second cluster, suprathreshold input leads to the last dimensional
reduction while the state is mapped directly onto the periodic orbit. 

In general, a periodic orbit is \textit{unstable} if, after a random
perturbation into its vicinity, one or more clusters are not re-synchronized
by simultaneous suprathreshold input but desynchronize due to simultaneous
subthreshold input. An unstable \textit{attractor} results if these
clusters are formed through synchronization in a region of state space
that is located \textit{remote} from the periodic orbit towards which
the state then converges. 

Although this is a discontinuous system with delayed interactions
such that there is no simple basin structure in the state space of
phases and signals, a three-dimensional cartoon of the basin structure
in a state space of phases may help to gain further insight about
how trajectories approach and retreat from an unstable attractor in
the presence of noise. 
\begin{figure*}
{\centering \resizebox*{12cm}{!}{\includegraphics{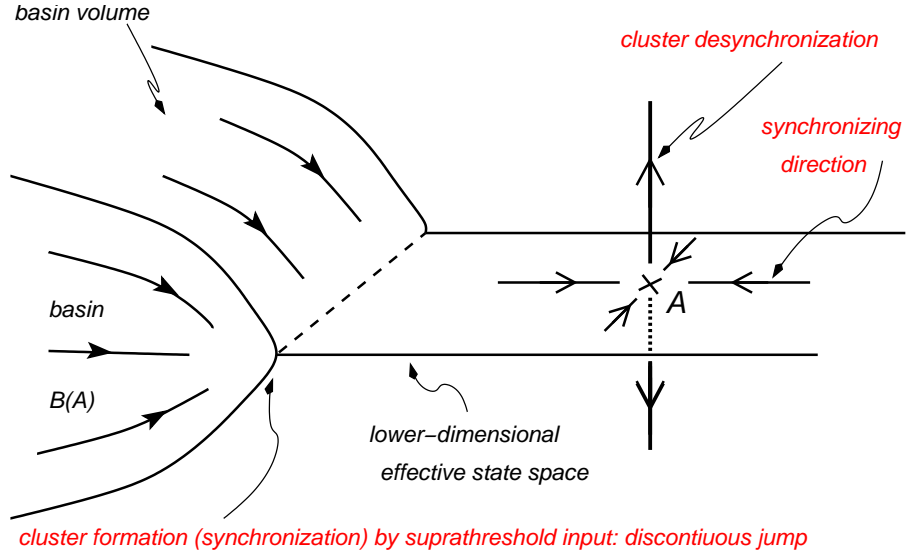}} \par}

\caption{Cartoon of the basin structure of an unstable attractor \protect\( A\protect \).
First, a positive basin volume \protect\( B(A)\protect \) of states
is mapped onto a lower-dimensional effective state space. This is
achieved by simultaneous suprathreshold input to a group of oscillators,
that synchronizes them to form one cluster. Because the attractor
is located remote from its own basin volume the same cluster may desynchronize
in response to small perturbations near the attractor, where incoming
pulses are no longer supra- but subthreshold and thus lead to a desynchronization.\label{Fig:cartoon}}
\end{figure*}
Figure \ref{Fig:cartoon} shows that the basin volume is contracted
by creating (at least) one cluster in a region of state space that
is remote from the attractor itself. In contrast, near the attractor,
the same cluster is unstable against a split-up of the phases of the
oscillators it contains. Basically, such an unstable attractor might
be viewed as an unstable periodic orbit with a remote basin attached
to its stable manifold that ensures the attractivity property. 

\begin{figure}
{\centering \resizebox*{6cm}{!}{\includegraphics{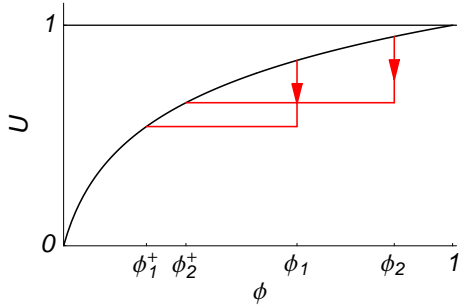}} \par}

\caption{{\small Simultaneous inhibitory input synchronizes due to the concavity
of \protect\( U\protect \). Contrary to networks of oscillators coupled
excitatorily, here the inhibitory interactions allow subthreshold
input only and lead to a decrease of phase differences}, {\small \protect\( \left| \phi _{2}(t^{+})-\phi _{1}(t^{+})\right| >\left| \phi _{2}(t)-\phi _{1}(t)\right| \protect \)
. Hence, these networks possess a mechanism for synchronization, but
there is no simple possibility of desynchronizing a cluster state.
\label{Fig:inhsyn}}}
\end{figure}

It is important to note that for inhibitory coupling we observe that
all attractors are stable: An intuitive explanation is that there
is only a mechanism of synchronization (Fig.~\ref{Fig:inhsyn}) due
to the concavity of \( U \) that contracts state space volume such
that all attractors with period-one dynamics are stable. It is instructive
to compare Fig.~\ref{Fig:inhsyn} for inhibition to Fig.~\ref{Fig:excdes}
for excitation that display how simultaneously incoming subthreshold
pulses affect phase differences.
 Since for inhibition all attractors are period-one states \cite{ErnstPRE,ErnstPRL},
we expect that the possibility of unstable attractors is excluded
for this kind of coupling. 

In order to further clarify the structure of state space of networks
of excitatorily coupled oscillators, we numerically determined the
basins of attraction of the three attractors displayed in Fig.~\ref{Fig:basin_N=3D6}a
in two-dimensional sections of state space. The example shown in Fig.~\ref{Fig:basin_N=3D6}b
reveals that attractors are surrounded by basins of attraction of
other attractors as predicted by the above analysis. Because of this
basin structure, noise induces repeated attractor switching among
unstable attractors. Starting from the orbit defined by (\ref{eq:syn})
the system may switch within sets of only six periodic orbit attractors
as is apparent from the basins shown in Fig.~\ref{Fig:basin_N=3D6}b.
However, in larger networks (cf.\ e.g.\ Fig.~\ref{Fig:noisy_phase_dynamics}d)
a cluster may split up in a combinatorial number of ways and exponentially
many periodic orbit attractors are present among which the system
may switch. The larger such networks are, the higher the flexibility
they exhibit in visiting different attractors and exploring state
space.

\begin{figure}
{\centering \includegraphics{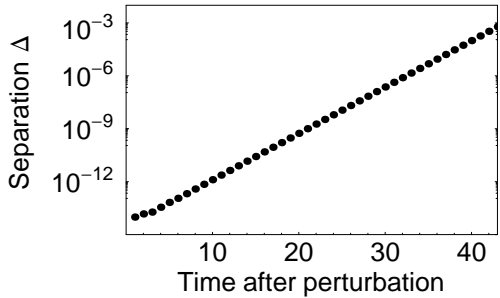} \par}

\caption{{\small Separation of two perturbations off an unstable attractor
into the same random direction for a large network (\protect\( N=100\protect \),
\protect\( \eps =0.2\protect \), \protect\( \tau =0.15\protect \)).
The separation \protect\( \Delta \protect \) grows exponentially
with the time after the perturbation, measured in number of firing
events of a reference oscillator. \label{Fig:separation_largeN}}}
\end{figure}

Until now, the analysis has focussed on a small network of \( N=6 \)
oscillators, for which certain periodic orbits have been demonstrated
to be unstable attractors. To study the desynchronization of clusters
also observed in larger network in greater detail, we numerically
determined the divergence of small random perturbations to an attractor
in a network of \( N=100 \) oscillators. As an example we chose two
perturbations \( \boldsymbol {\delta _{1}}=\sigma _{1}\boldsymbol {\delta ^{*}} \)
and \( \boldsymbol {\delta _{2}}=\sigma _{2}\boldsymbol {\delta ^{*}} \)
into the same random direction \( \boldsymbol {\delta ^{*}}\in [0,1]^{N} \)
where \( \sigma _{1}=10^{-12} \) and \( \sigma _{2}=\sigma _{1}+10^{-14} \).
Figure \ref{Fig:separation_largeN} shows that the separation \( \Delta :=\max _{i}|\delta _{1,i}-\delta _{2,i}| \)
between the two perturbed trajectories exponentially increases with
time. This indicates that also for large networks, desynchronization
is due to a linear instability. Let us remark that, due to the spiltting-up
of clusters by a general perturbation (cf.\ Eqs.\ (\ref{eq:PO'})
and (\ref{eq:C_prime})), two perturbations into independent directions
might first lead to an additional discontinuous separation, followed
by an exponential expansion.

\begin{figure}
{\centering \includegraphics{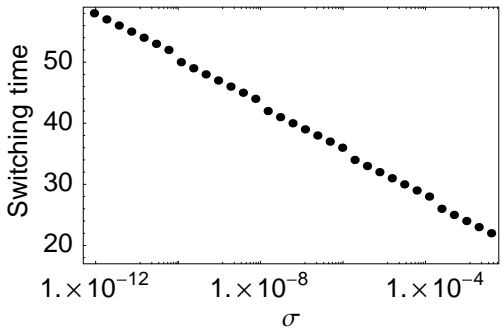} \par}

\caption{{\small Time of switching between two attractors depending on the
perturbation strength \protect\( \sigma \protect \) for a large network
(\protect\( N=100\protect \), \protect\( \eps =0.2\protect \), \protect\( \tau =0.15\protect \)).
The switching time increases exponentially with decreasing perturbation
strength. The discrete time axis (number of firing events of a reference
oscillator) leads to a regular stepping that slightly disrupts the
exponential trend. \label{Fig:switch_largeN}}}
\end{figure}

As a second quantity, that characterizes the dynamics of switching,
we consider the time needed by the system to switch from an unstable
attractor towards a different attractor after a random perturbation
of magnitude \( \sigma  \). The perturbations applied to the attractor
are random phase vectors, drawn from a uniform distribution on \( [0,\, \sigma ]^{N} \),
with \( \sigma \in [10^{-12},10^{-2}] \). As displayed in Fig.\ \ref{Fig:switch_largeN},
for sufficiently small \( \sigma  \), this switching time clearly
increases exponentially with decreasing \( \sigma  \). In particular,
this indicates that the attractors found are indeed unstable and do
not possess small contracting open neighborhoods. Furthermore, it
confirms our observations that the time of switching is mainly determined
by the time of divergence from the original unstable attractor. Thus,
in the presence of external noise, we expect a similar monotonic increase
of an approximate switching time with the amplitude of the noise.

Taken together, this indicates that there exist also unstable attractors
in larger networks that are enclosed by basins of other attractors.
Thus, these results strongly support the hypothesis, that the switching
found in large networks in the presence of noise (cf.\ Fig.\ \ref{Fig:noisy_phase_dynamics}),
is also due to unstable attractors. It is important to note that,
without a perturbation, numerical noise does not induce a divergence
of trajectories from attractors which are unstable: Synchronization
occurs by simultaneously resetting the phases of two or more oscillators
to zero (cf.\ Fig.\ \ref{Fig:excsyn}). Due to the global homogeneous
coupling, all signals simultaneously received by these oscillators
are of the same size and are exactly synchronous numerically (see
Sec.\ \ref{sec:Mirollo_Strogatz_model}). Thus, although the phase-advance
computed for such signals is influenced by numerical round-off errors,
such errors will be identical for phases of synchronized oscillators
and hence not induce a numerical desynchronization.

\section{Prevalence and persistence}

\label{sec:prevalence}

The preceeding analysis demonstrates the existence of unstable attractors.
For excitatory coupling, these unstable attractors coexist with stable
attractors. For a cluster-state attractor to be stable, all clusters
necessarily receive suprathreshold input once per period that re-synchronizes
possibly split-up oscillators. In general, a stable attractor has
a contracting neighborhood in state space from which perturbed trajectories
return to the attractor. If an attractor is unstable there are trajectories
arbitrarily close to the attractor, which diverge from it, such that
unstable attractors do not exhibit a contracting neighborhood. In
other words, an unstable attractor has to be located at the boundary
of its own basin. Therefore, the intuitive expectation is that parameters
of the system have to be precisely specified in order to keep a periodic
orbit simultaneously attracting and unstable. This leads to the question
whether the physical parameters of the system, in particular \( \eps  \),
\( \tau  \), and \( N \), need to be precisely tuned to obtain unstable
attractors. 

\begin{figure}
{\centering {\small \resizebox*{7cm}{!}{\includegraphics{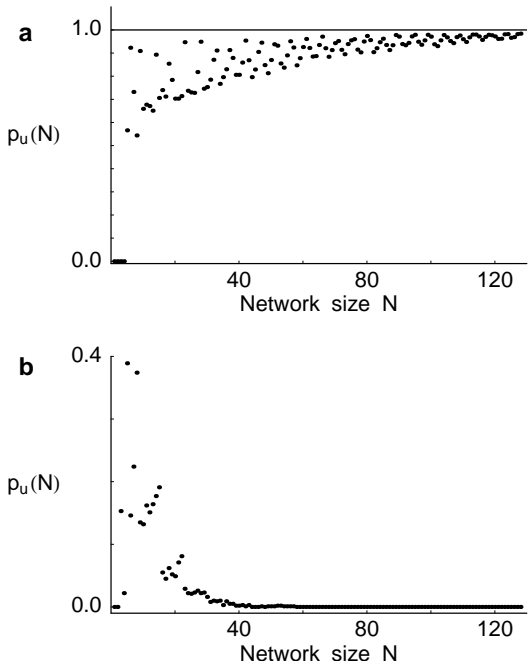}} } \par}

\caption{Prevalence of unstable attractors in large networks. (a) Unstable
attractors prevail for large networks for certain parameters (\protect\( \eps =0.2\protect \),
\protect\( \tau =0.15)\protect \), but (b) are not important in large
networks for other parameters (\protect\( \eps =0.2\protect \), \protect\( \tau =0.25\protect \)).
The fraction \protect\( p_{\mathrm{u}}(N)\protect \) was estimated
for every \protect\( N\leq 128\protect \) from \protect\( 1000\protect \)
random initial phase vectors, drawn from the uniform distribution
on \protect\( [0,1]^{N}\protect \). \label{Fig:prevalence_N}}
\end{figure}

To answer the question, how common unstable attractors actually are,
we numerically estimated the fraction \( p_{\mathrm{u}}(N) \) of
state space occupied by basins of unstable attractors. To obtain this
estimate, we initialized the system with 1000 random initial phase
vectors, drawn from the uniform distribution on \( [0,1]^{N} \).
Whenever a period-one orbit was reached, we applied one random phase
perturbation \( \boldsymbol {\delta } \) drawn from the uniform distribution
on \( [0,\sigma ]^{N} \)where we chose \( \sigma =10^{-6} \), a
value well below all scales that are determined by the model parameters,
in particular \( \sigma \ll \eps /N \) for network sizes up to \( N\approx 10^{2} \).
If perturbed trajectories did not return to the original attractor,
it was counted unstable. If no period-one orbit was reached from a
random initial state but, e.g., orbits of higher period, these were
not tested for stability. Thus, the numerical method used estimates
a lower bound on \( p_{\mathrm{u}}(N) \). As an example, Fig.~\ref{Fig:prevalence_N}a
displays such an estimate of \( p_{\mathrm{u}}(N) \) for \( \varepsilon =0.2 \)
and \( \tau =0.15 \).

\begin{figure}
{\centering {\small \resizebox*{7cm}{!}{\includegraphics{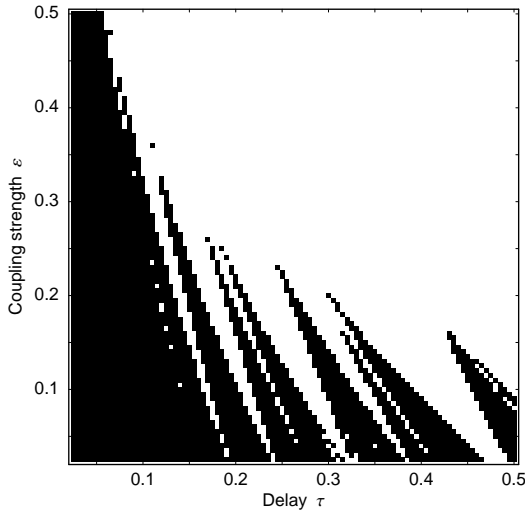}} } \par}

\caption{Unstable attractors persist in a wide region of parameter space in
large networks (\protect\( N=100\protect \)). Parameters with \protect\( p_{\mathrm{u}}(100)>0.5\protect \)
are marked in black; here \protect\( p_{\mathrm{u}}(100)\protect \)
was estimated from \protect\( 100\protect \) random initial phase
vectors, drawn from the uniform distribution on \protect\( [0,1]^{N}\protect \),
for every set of parameters (resolution \protect\( \Delta \tau =\Delta \eps =0.005\protect \)).\label{Fig:prevalence_eps_tau}}
\end{figure}
While for these parameters unstable attractors are absent if networks
are too small (here \( N\leq 4 \)) and coexist with stable attractors
in larger networks, the fraction approaches one for \( N\gg 1 \).
Other parameters (\( \eps =0.2 \), \( \tau =0.25 \)) yield a different
dependence on the network size \( N \). Once again, unstable attractors
arise only if the network size is not too small (\( N\geq 3) \).
Yet, the fraction \( p_{\mathrm{u}}(N) \) is only substantial for
moderate network sizes near \( N\approx 10 \) and approaches zero
for large \( N \). More generally, we observed that unstable attractors
are absent in small networks (cf.\ \cite{ErnstPRE,ErnstPRL} for the
case \( N=2 \)) and \( p_{\mathrm{u}}(N) \) approaches either zero
or one in large networks, depending on the parameters. For networks
of \( N=100 \) oscillators, Fig.~\ref{Fig:prevalence_eps_tau} shows
the region of parameter space in which unstable attractors prevail
(\( p_{\mathrm{u}}(100)>0.5 \), estimated from 100 random initial
phase vectors). As this region covers a substantial part of parameter
space, precise parameter tuning is not needed to obtain unstable attractors.
Furthermore we find the same qualitative behavior independent of the
detailed form of \( U(\phi ) \). This indicates that the occurrence
of unstable attractors is a robust collective phenomenon in this model
class of networks of excitatorily pulse-coupled oscillators.

It is instructive to note that, on theoretical grounds, a period-one
orbit is stable only if the clusters have a difference in firing times
of \( \Delta t=\tau  \), such that its period is an integer multiple
of \( \tau  \). The pulse sent by every single cluster then leads
to a suprathreshold input to the following cluster, that in response
to this input sends a pulse. If these conditions are not satisfied
for all clusters of a period-one orbit, this orbit is unstable. In
particular, it is unstable against a split-up of (at least) one cluster.
The same orbit may also be attracting if such a cluster is formed
in a region of state space that is located remote from the attractor
as exemplified by the analysis in section \ref{sec:unstable_attractors}. 

Whereas the analysis presented above gives some insights into why
unstable attractors exist and demonstrates that they prevail under
variation of parameters, the precise reasons for their prevalence
await discovery in future studies.

\section{Conclusions and Discussion}

\label{sec:conclusions}

The occurrence of unstable attractors \textit{per se} is an intriguing
phenomenon because it contradicts the common intuition about the stable
nature of attracting invariant sets in dynamical systems. Our results
suggest, that there are systems of pulse-coupled units in which unstable
attractors may be the rule rather than the exception.

Unstable attractors persist under various classes of structural modifications.
For instance, preliminary studies on networks with randomly diluted
connectivity suggest, that a symmetric, all-to-all connectivity is
not required \cite{Zumdieck}. In addition, unstable attractors also
arise naturally in networks of inhibitorily coupled oscillators \cite{Denker},
if a lower threshold is introduced and the function \( U \) is taken
to be convex down, \( U''(\phi )>0 \), in a certain range of phase
values, a model variant motivated by experiments in certain biological
neural systems \cite{Johnston}.

Moreover, it is expected that every system obtained by a sufficiently
small structural perturbation from the one considered here will exhibit
a similar set of saddle periodic orbits, because linearly unstable
states can generally not be stabilized by such a perturbation. Although,
in general, these orbits may no longer be attracting, their dynamical
consequences are expected to persist. In particular, a switching along
heteroclinic connections may occur in the presence of noisy or deterministic,
time-varying signals. As in the original system, the sequence of states
reached may be determined by the directions into which such a signal
guides the trajectory. By increasing and decreasing the strength of
this signal, the time-scale of switching may be decreased and increased,
respectively, due to the linear instability. 

Furthermore, switching among unstable states also occurs in systems
of continuously phase-coupled oscillators \cite{HanselPRE,Kori} that
can be obtained from pulse-coupled oscillators in a certain limit
of weak coupling \cite{Kuramoto}. In particular, Hansel, Mato, and
Meunier \cite{HanselPRE} show that a system of phase-coupled oscillators
may switch back and forth among pairs of two-cluster states. Working
in the limit of infinitely fast response, i.e.\ discontinuous phase
jumps, we have demonstrated that far more complicated switching transitions
may occur in large networks if the oscillators are pulse-coupled.

Unstable attractors add a high degree of flexibility to a system allowing,
e.g., switching from one attractor towards a set of other attractors.
This may be utilized for specific functions. If, for instance, the
convergence of the state of the system towards an attractor has a
functional role, such as the solution of a computational task \cite{Hertz,Hopfield,Maass},
the flexibility induced by unstable attractors can provide the system
with a unique advantage: If the attractor is stable, it will be hard
to leave it after convergence, e.g. the completion of a task. With
an unstable attractor, however, a small perturbation is sufficient
to leave the attractor after convergence and proceed with the next
task. This dynamical flexibility might be used efficiently for the
design of artificial systems and be highly advantageous to the computational
capability of natural systems like neuronal networks. Interestingly,
it has recently been shown that certain models of neural networks
are capable of dynamically encoding information as trajectories near
heteroclinic connections \cite{Rabinovich}. 

In order to fully understand the capabilities of systems exhibiting
unstable attractors or unstable periodic orbits linked by heteroclinic
connections and to learn to design such systems for specific functions
it will be of major importance to further analyze the requirements
for the occurrence of unstable attractors in dynamical systems and
the factors which shape their basins. Our results indicate that networks
of pulse-coupled oscillators may be a promising starting point for
such investigations.

\begin{acknowledgments}
We thank A.\ Aertsen, M.\ Diesmann, U.\ Ernst, D.\ Hansel, K.\ Kaneko,
K.\ Pawelzik and C.\ v.\ Vreeswijk for useful discussions.
\end{acknowledgments}

\end{document}